\documentclass[reprint,superscriptaddress]{revtex4-1}
\usepackage{amsmath}
\usepackage{amssymb}
\usepackage{cancel}
\usepackage{amsfonts}
\usepackage[T1]{fontenc}
\usepackage{bm}
\usepackage{color}
\usepackage{graphicx}
\DeclareSymbolFont{operators}{OT1}{cmr}{m}{n}
\DeclareSymbolFont{letters}{OML}{cmm}{m}{it}
\DeclareSymbolFont{symbols}{OMS}{cmsy}{m}{n}
\DeclareSymbolFont{largesymbols}{OMX}{cmex}{m}{n}

\usepackage{hyperref}
\hypersetup{
    bookmarks=true,         
    unicode=false,          
    pdftoolbar=true,        
    pdfmenubar=true,        
    pdffitwindow=false,     
    pdfstartview={FitH},    
    pdfsubject={Plasma fluid theory},   
    pdfproducer={GNU Make}, 
    pdfkeywords={keyword1} {key2} {key3}, 
    pdfnewwindow=true,      
    colorlinks=true,        
    linkcolor=blue,      
    citecolor=blue,         
    filecolor=blue,      
    urlcolor=blue           
}
\usepackage{array}

\newcommand{\mhh}[1]{{\bf \bar{H}}_{(#1)}}
\newcommand{\mg}[1]{{\bf \bar{G}}_{(#1)}}

\newcommand{\1}{\hat{\bm{1}}}

\makeatother

\begin{document}
\title{Nonlinear Resistivity for Magnetohydrodynamical Models}

\author{M. Lingam}
\email{manasvi@seas.harvard.edu}
\affiliation{Department of Astrophysical Sciences, Princeton University, Princeton, NJ 08544, USA}
\affiliation{Harvard John A. Paulson School of Engineering and Applied Sciences, Harvard University, Cambridge, MA 02138, USA}
\affiliation{Harvard-Smithsonian Center for Astrophysics, The Institute for Theory and Computation, Cambridge, MA 02138, USA}
\author{E. Hirvijoki}
\affiliation{Princeton Plasma Physics Laboratory, Princeton, New Jersey 08543, USA}
\author{D. Pfefferl\'e}
\affiliation{Princeton Plasma Physics Laboratory, Princeton, New Jersey 08543, USA}
\author{L. Comisso}
\affiliation{Department of Astrophysical Sciences, Princeton University, Princeton, NJ 08544, USA}
\affiliation{Princeton Plasma Physics Laboratory, Princeton, New Jersey 08543, USA}
\author{A. Bhattacharjee}
\affiliation{Department of Astrophysical Sciences, Princeton University, Princeton, NJ 08544, USA}
\affiliation{Princeton Plasma Physics Laboratory, Princeton, New Jersey 08543, USA}

\begin{abstract}
A new formulation of the plasma resistivity that arises from the collisional momentum-transfer rate between electrons and ions is presented. The resistivity computed herein is shown to depend not only on the temperature and density but also on all other polynomial velocity-space moments of the distribution function, such as the pressure tensor and heat flux vector. The exact expression for the collisional momentum-transfer rate is determined, and is used to formulate the nonlinear anisotropic resistivity. The new formalism recovers the Spitzer resistivity, as well as the concept of thermal force if the heat flux is assumed to be proportional to a temperature gradient. Furthermore, if the pressure tensor is related to viscous stress, the latter enters the expression for the resistivity. The relative importance of the nonlinear term(s) with respect to the well-established electron inertia and Hall terms is also examined. The subtle implications of the nonlinear resistivity, and its dependence on the fluid variables, are discussed in the context of magnetized plasma environments and phenomena such as magnetic reconnection.
\end{abstract}

\pacs{put Pacs here}%
 
\maketitle

\section{Introduction}
The ubiquity and importance of fluid models in plasma physics stems
from their versatility and simplicity
\citep{Spit56,KT73,KR95,HW04,GP04,Fre14}. Amongst all such models,
ideal magnetohydrodynamics (MHD) remains the best known. Besides ideal
MHD, other variants of MHD such as Hall MHD, electron MHD and extended
MHD have also been widely studied and utilized. These models are
derived by taking the moments of the Boltzmann equation, and then
carrying out formal expansions involving a small parameter. The
Boltzmann equation is endowed with a collision operator, whose effects
are manifested in the form of dissipative terms.

One of these, the electrical resistivity, plays a particularly
important role, as it gives rise to new phenomena and behaviour that
are otherwise inaccessible in its absence. For example, extensively
studied processes such as turbulence, magnetic reconnection, and
dynamos represent striking examples of the importance of resistivity
in space and astrophysical plasmas
\citep{Spit56,Dung58,KR95,Gomb98,BG05,Kul05,Pri14}. Yet, an important
(and implicit) feature is inherent in many of these studies, namely,
the resistivity depends only on the temperature and density; the dependence on the latter quantity is rather weak (logarithmic). As the
resistivity ultimately stems from the collisional electron-ion
momentum-transfer rate, it has, in fact, a nonlinear dependence on all
of the fluid variables.

Hence, the primary objective of this paper is to derive an expression
for resistivity that correctly accounts for the nonlinear nature of
the Landau collision operator. Our work adopts the methodology
introduced in \citep{HLPC16} and the recent results from
\citep{PHL2017}, enabling us to express the collisional
momentum-transfer rate between the species $s$ and
$s'$ without employing any assumptions. The collisional momentum-transfer rate computed herein exhibits a linear dependency on the velocity-space moments of the distribution functions such as the pressure tensor and the heat flux. The so-called thermal force is thus recovered in the fluid dynamical limit where the heat flux is proportional to a temperature gradient. Furthermore, if
pressure anisotropy is proportional to the viscous stress, the
viscosity will enter the expression for the collisional
momentum-transfer rate.

The outline of the paper is as follows. In Sec. \ref{SecSummary}, we
briefly summarize the standard approach for deriving the extended MHD
Ohm's law. In Sec. \ref{SecResTensor}, we present the general
expression for the resistivity tensor, and explore its various limits. In
Sec. \ref{SecMaxwell}, the physical consequences arising from current
and pressure (anisotropy) dependent resistivity are explored. Finally,
we summarize our findings and implications in Sec. \ref{SecConc}.

\section{The generalized Ohm's law} 
\label{SecSummary} 
The first step in deriving fluid models entails taking the moments of
the Boltzmann equation \citep{HW04}. The zeroth moment leads to the
continuity equation of each species $s$, whilst the first moment leads
to the corresponding momentum equation. The latter is given by
\begin{multline} \label{Mom}
\frac{\partial \left(n_s m_s \bm{V}_{s}\right)}{\partial t} + \nabla\cdot\left(\mathbb{P}_{s}+m_s n_s \bm{V}_{s} \bm{V}_{s}\right)\\=q_s n_s \left(\bm{E}+\bm{V}_{s}\times\bm{B}\right)+\bm{F}_{s s'},
\end{multline}
where $n_s$ is the number density, whilst $m_s$ and $q_s$ are the mass
and the charge respectively. $\bm{E}$ is the electric field, $\bm{B}$
is the magnetic field, $\mathbb{P}_s$ is the pressure tensor, and
$\bm{F}_{s s'}$ is the collisional momentum transfer rate between the
species $s$ and $s'$. The second moment of the kinetic equations leads
to the equation for the pressure tensor, which we shall not present
here explicitly. In addition to these moments, the Maxwell equations
must also be taken into account.

The next stage follows by specializing to an electron-ion plasma and
reducing the Maxwell equations to the pre-Maxwell equations. This is
done by demanding that the characteristic thermal and (electromagnetic
wave) phase velocities are non-relativistic, which enables the neglect
of the displacement current. Secondly, the domain is restricted to
characteristic frequencies $\left(\omega\right)$ and length scales
$\left(L\right)$ that obey $\omega \ll \omega_{pe}$ and $L \gg
\lambda_D$, where $\omega_{pe}$ is the electron plasma frequency and
$\lambda_D$ is the Debye length. This ensures that the model is
quasineutral, viz. $n_i = n = n_e$. We refer the reader to Sec. 2.3.3
of \citet{Fre14} (see also \citet{Fitz14}) for further details.

Following the standard prescription for one-fluid variables, the
center-of-mass velocity $\bm{V}=\left(m_i \bm{V}_i + m_e
  \bm{V}_e\right)/\left(m_i + m_e\right)$ and the current density
$\bm{J} = n e \left(\bm{V}_i-\bm{V}_e\right) = \mu_0^{-1} \nabla
\times \bm{B}$ are introduced. For an ion-electron plasma, the mass
ratio $\delta = m_e/m_i$ is a legitimate `small' parameter for
carrying out asymptotic expansions. The ion and electron flow
velocities can be expressed in terms of the one-fluid variables:
\begin{equation}
\bm{V}_i=\bm{V}+\frac{\delta}{1+\delta}\frac{\bm{J}}{ne},\qquad
\bm{V}_e=\bm{V}-\frac{1}{1+\delta}\frac{\bm{J}}{ne}.
\end{equation}
Upon adding and subtracting the two species equations, using the
continuity equations, and consistently dropping the terms of ${\cal
  O}(\delta)$ or higher, the final set of equations can be duly
obtained. They comprise of the continuity equation for $n$, the
momentum equation
\begin{equation} \label{XMHDMom}
\rho \left(\frac{\partial \bm{V}}{\partial t} + \bm{V} \cdot \nabla \bm{V}\right) = - \nabla \cdot \mathbb{P}\, +\, \bm{J} \times \bm{B} - \frac{m_e}{e} \bm{J} \cdot \nabla \left(\frac{\bm{J}}{ne}\right),
\end{equation}
and the generalized Ohm's law
\begin{multline} \label{XMHDOhm}
\bm{E}+\bm{V}\times\bm{B} = \frac{\bm{F}_{ei}}{ne} + \frac{\bm{J} \times \bm{B} - \nabla \cdot \mathbb{P}_e}{ne}
\\
+ \frac{m_e}{n e^2} \left[\frac{\partial \bm{J}}{\partial t} + \nabla \cdot \left(\bm{V} \bm{J} + \bm{J} \bm{V} - \frac{\bm{J} \bm{J}}{ne}\right)\right],
\end{multline}
where $\mathbb{P} = \mathbb{P}_i + \mathbb{P}_e$ is the total pressure
tensor, and $\rho = n \left(m_e + m_i\right)$ denotes the mass
density. Both (\ref{XMHDMom}) and (\ref{XMHDOhm}) are fully consistent
with other treatments in the literature; see for
e.g. \citet{Dung58,Gomb98,GP04,BG05}. Some textbooks and papers omit
the last terms on the RHS of (\ref{XMHDMom}) and (\ref{XMHDOhm}), as
they use a different definition for the pressure tensors, and thereby
absorb these terms into the latter accordingly. In order to close the
system, the evolution equations for the pressure tensors (of each
species) must also be taken into account \citep{PHL2017}.

In this paper, we focus on the generalized Ohm's law (\ref{XMHDOhm}),
with a particular emphasis on determining the contributions arising
from the collisional electron-ion momentum transfer rate. We will now
proceed to show that our results diverge from the standard results
owing to the differences stemming from $\bm{F}_{ei}$. Given that the
dynamical equations for the species-wise pressure involve terms of the
form ${\bf V}_s \cdot {\bf E}$ \citep{KT73,GP04}, the expression for
$\bm{F}_{ei}$ also has ramifications for the evolution of the
pressure.

\section{Collisional momentum transfer rate and the resistivity tensor}
\label{SecResTensor}
In this Section, we discuss the expression for collisional
momentum-transfer rate $\bm{F}_{ss'}$, and use it to determine
$\bm{F}_{ei}/(ne)$ while assuming the MHD ordering. This will lead us to an exact expression for the resistivity tensor.

\subsection{Conventional approach}
Before presenting the results from our approach, we shall summarize some of the salient results obtained through standard approaches.

For the sake of simplicity, fluid models often express the term $\bm{F}_{ei}/(ne)$ using the lowest order result
\begin{equation}
\frac{\bm{F}_{ei}}{ne}=\eta_0 \bm{J}, \qquad \eta_0 = \frac{4}{3} \frac{\sqrt{2\pi m_e}e^2\ln\Lambda}{(4\pi\varepsilon_0)^2 T_e^{3/2}},   \end{equation}
where $\eta_0$ denotes the conventional (Spitzer) resistivity \citep{Spit56,HM92}
and $\ln\Lambda$ is the Coulomb logarithm. Further corrections to $\bm{F}_{ei}$ are typically obtained by means of Braginskii's transport theory \citep{Brag65} which, in the limit of strong magnetic field, leads to the expression
\begin{equation}
\label{eq:braginskii}
\frac{\bm{F}_{ei}}{ne}=\eta_0\left[0.51\frac{\bm{B}\bm{B}}{B^2}+\left(\mathbb{I}-\frac{\bm{B}\bm{B}}{B^2}\right)\right]\cdot\bm{J}+\frac{\bm{R}_{T}}{ne},
\end{equation}
where $\bm{R}_T$ is the so-called thermal force 
\begin{equation} \label{ThermFo}
\frac{\bm{R}_T}{ne}=-0.71\frac{\bm{B} \bm{B}}{eB^2}\cdot\nabla T_e-\frac{3\eta_0}{2}\frac{n}{ B}\frac{\bm{B}}{B}\times\nabla T_e.
\end{equation}
Although Braginskii's method for strongly magnetized plasmas closely follows Chapman-Enskog theory \citep{CC70}, the presence of two species and the magnetic force significantly complicate the procedure: 
\begin{itemize}
\item The magnetic force and collisions are ordered to occur on the fastest time scale. This is in contrast to the standard Chapman-Enskog approach for neutral fluids, where only the collisions are ordered in such a manner.
\item Because of the magnetic force, the velocity and its derivatives in the Vlasov operators must be written in terms of the so-called random-velocity variables $\bm{w}_s=\bm{v}-\bm{V}_s$, to guarantee that a Maxwellian with a flow velocity is a null-eigensolution of the lowest order Vlasov operator (the magnetic term).
\item Due to the presence of ions, the electron-ion collision operator must be approximated and split into two terms so that, in the lowest order kinetic equation, the Maxwellian with electron flow velocity remains a null-eigensolution.
\item The change-of-variables with respect to the velocity coordinates leads to time derivatives of the respective flow velocities appearing in the equations for the first corrections to the distribution functions. These must be replaced by using the momentum equations that include the collisional momentum-transfer rate. In addition, the contribution from the electron-ion collision that was left out from the lowest order kinetic equation, must be included and approximated further.
\end{itemize}
Hence, a procedure that can be carried out quite rigorously for neutral fluids in the limit of strong collisionality becomes more complex for multi-component plasmas due to the various approximations required even prior to obtaining the first-order correction equations.

It is observed that the so-called variational solution to the integral equations, from which the transport coefficients are solved for in both the classic Chapman-Enskog and Braginskii methods, involves a polynomial expansion of the corrections to the distribution functions, regardless of the specific approximations introduced to derive the integral equations. It also happens that the collisional momentum-transfer rate can be evaluated exactly in terms of coefficients that are orthogonal-polynomial velocity-space moments of the distribution functions. In other words, the exact expression for the collisional momentum-transfer rate can thus be expressed in terms of the fluxes that the classical transport theory is solving for (independently of the explicit expressions for these fluxes).

\subsection{Exact expression for collisional momentum transfer}
To improve upon the previous results, a more accurate expression for
the collisional momentum-transfer rate would be necessary. This can be duly obtained by using spectral expansions for the distribution functions in terms of multi-index Hermite polynomials, in which case the collisional
momentum transfer rate can be evaluated exactly
\citep{PHL2017}. Defining the function
\begin{align}
  \Phi(z) &= \int_{\mathbb{R}^3}\frac{d\bm{x}}{|\bm{x}|}\frac{e^{-(\bm{x}-\bm{z})^2}}{\pi^{3/2}}=\frac{\text{erf}(z)}{z},
\end{align}
the coefficients $\mu_{ss'}=1/m_s+1/m_{s'}$,
$\bar{c}_{ss'}=n_sn_{s'}\ln\Lambda (e_se_{s'})^2/4\pi\varepsilon_0^2$,
$\sigma_s^2=T_s/m_s$, $\Sigma^2_{ss'}=\sigma_s^2+\sigma_{s'}^2$, and
$\bm{\Delta}_{ss'}=(\bm{V}_s-\bm{V}_{s'})/\sqrt{2}\Sigma_{ss'}$, the
collisional momentum-transfer rate is given by
\begin{multline}
\label{eq:momentum-transfer}
\bm{F}_{ss'}
=\frac{\bar{c}_{ss'}\mu_{ss'}}{2\Sigma_{ss'}^2}\partial_{\bm{\Delta}_{ss'}}\sum_{i,j=0}^\infty \frac{(-1)^j}{i!j!}
\frac{\bm{c}_{s(i)}\bm{c}_{s'(j)}}{(\sqrt{2}\Sigma_{ss'})^{i+j}}\\\partial_{\bm{\Delta}_{ss'}}^{(i)}\partial_{\bm{\Delta}_{ss'}}^{(j)}
 \Phi(\Delta_{ss'}), 
\end{multline}
where $\Delta_{ss'} = |\bm{\Delta}_{ss'}|$. Since the coefficient vector is antisymmetric $\bm{\Delta}_{ss'}=-\bm{\Delta}_{s's}$ while its magnitude remains symmetric $\Delta_{ss'}=\Delta_{s's}$, and the scalar coefficients $\bar{c}_{ss'}$, $\mu_{ss'}$, and $\Sigma_{ss'}$ are symmetric with respect to exchanging the species indices, we find
\begin{multline}
\bm{F}_{ss'}
=-\frac{\bar{c}_{s's}\mu_{s's}}{2\Sigma_{s's}^2}\partial_{\bm{\Delta}_{s's}}\sum_{i,j=0}^\infty \frac{(-1)^i}{i!j!}
\frac{\bm{c}_{s(i)}\bm{c}_{s'(j)}}{(\sqrt{2}\Sigma_{s's})^{i+j}}\\\partial_{\bm{\Delta}_{s's}}^{(i)}\partial_{\bm{\Delta}_{s's}}^{(j)}
 \Phi(\Delta_{s's})=-\bm{F}_{s's}, 
\end{multline}
which demonstrates that the momentum-transfer rate is explicitly antisymmetric (required for the conservation of momentum) to \emph{all orders} in $\bm{c}_{s(i)}$ and $\bm{c}_{s'(j)}$. The spectral expansion coefficients $\bm{c}_{s(i)}$ are multi-index Hermite polynomial moments of the distribution functions, 
\begin{align}\label{eq:distro_hermites}
\bm{c}_{s(j)}\equiv \int_{\mathbb{R}^3} d\bm{v} \frac{f_s(\bm{v})}{n_s}\mhh{j}(\bm{v}-\bm{V}_s;\sigma_s^2),
\end{align}
with the Hermite polynomial $\mhh{k}$ defined as 
\begin{align}\label{eq:hermite_def}
\mhh{k}(\bm{x};\sigma^2)\mathcal{N}_{\sigma^2}(\bm{x}) =(-\sigma^2\partial_{\bm{x}})^{(k)} \mathcal{N}_{\sigma^2}(\bm{x}),
\end{align}
that is generated by means of the three-dimensional Normal distribution
\begin{equation}\label{eq:normal_distro}
\mathcal{N}_{\sigma^2}(\bm{x})\equiv \frac{e^{-x^2/2\sigma^2}}{(2\pi)^{3/2}\sigma^3}.
\end{equation}

We wish to stress that the above result is \emph{exact}: the Landau collision operator has not been approximated to any degree. We observe that $\bm{F}_{ss'}$ implicitly includes the coefficients $\bm{c}_{s(i)}$, and will consequently depend on quantities such as the current, density, temperature, pressure anisotropy, heat flux, etc. Next, we shall describe how our result (\ref{eq:momentum-transfer}) is consistent with those obtained from the Chapman-Enskog and Braginskii approaches, for e.g. the thermal force (\ref{ThermFo}).

\subsection{Connection to conventional transport theory}
In the Braginskii and Chapman-Enskog formulations, the distribution functions are given by $f_s=f_{s0}(1+\Psi_{s})$, with the dominant contribution being a Maxwellian
\begin{equation}
f_{s0}(\bm{v})=n_s{\cal N}_{\sigma_s^2}(\bm{w}_s).
\end{equation}
where $\bm{w}_s=\bm{v}-\bm{V}_s$. The correction to the Maxwellian distribution function is not allowed to carry density, momentum, or energy, and thus $\Psi_{s}$ has a generic form that consists of a scalar, a random-velocity vector contracted with a vector, and a traceless rank-2 tensor of the random velocity that is contracted with another rank-2 tensor. If expanded in Laguerre polynomials, the expression for $\Psi_s$ can be written as 
\begin{multline}
\label{eq:CE-ansatz}
\Psi_{s}(\bm{v})=\sum_{m=2}^{\infty}\alpha_s^{m}L_{m}^{(1/2)}(y_s)+\bm{w}_s\cdot\sum_{m=1}^{\infty}\bm{\beta}_s^{m}L_m^{(3/2)}(y_s)\\+\left(\bm{w}_s\bm{w}_s-\frac{w_s^2}{3}\mathbb{I}\right):\sum_{m=0}^{\infty}\bm{\gamma}_s^mL_m^{(5/2)}(y_s),
\end{multline}
where $y_s=w_s^2/(2\sigma_s^2)$. It is then the task of the conventional transport theory to determine the coefficients $\alpha_{s}^m,\bm{\beta}_s^{m},\bm{\gamma}_s^m$ by using the ansatz (\ref{eq:CE-ansatz}) to convert the integral equations in the Chapman-Enskog theory into a system of algebraic equations. Furthermore, from the definitions for heat flux and viscosity, one concludes that the lowest order expression for $\Psi_s$ is given by
\begin{equation} \label{Psexp}
\Psi_s=\left(\frac{w_s^2}{5\sigma_s^2}-1\right)\bm{w}\cdot\frac{\bm{q}^{[0]}_s}{\sigma^2_s p_s} + \bm{w}_s\bm{w}_s:\frac{\bm{\pi}^{[0]}_s}{2\sigma^2_s p_s},
\end{equation}
regardless of the explicit expressions for the lowest order approximations of heat flux $\bm{q}^{[0]}_s$ and the viscosity $\bm{\pi}^{[0]}_s$  \citep[see pp. 243-245]{Brag65}. It is more convenient to write the lowest order expression for $\Psi_s$ in terms of the so-called covariant Hermite polynomials.
\begin{equation}
  \label{eq:covariant_gen}
  \mg{k}(\bm{x};\sigma^2)\mathcal{N}_{\sigma^2}(\bm{x}) =(-\partial_{\bm{x}})^{(k)}   \mathcal{N}_{\sigma^2}(\bm{x}), 
\end{equation}
in which case one arrives at
\begin{equation}
\Psi_s=\delta^{ij}\mg{3}^{ijk}(\bm{w}_s;\sigma_s^2)\frac{\sigma_s^2 q^{k[0]}_s}{5p_s} + \mg{2}^{ij}(\bm{w}_s;\sigma_s^2):\frac{\sigma_s^2\pi^{ij[0]}_s}{2 p_s}.
\end{equation}
This enables us to find the expansion coefficients $\bm{c}_{s(i)}$ which are required for computing $\bm{F}_{ss'}$. Using the above ansatz for $\Psi_s$, we find
\begin{align}
c_{s(0)}&=1\\
\bm{c}^{ij}_{s(2)}&=\sigma_s^2\frac{\pi_s^{ij}}{p_s}\\
\bm{c}^{ijk}_{s(3)}&=\frac{2\sigma_s^2}{5p_s}(q^{i[0]}_s\delta^{jk}+q^{j[0]}_s\delta^{ik}+q^{k[0]}_s\delta^{ij})
\end{align}
while the rest of the coefficients vanish identically.

A couple of observations are in order at this juncture. Firstly, there exists a correspondence between the Laguerre polynomial expansion employed by Braginskii, and contractions of higher-order Hermite polynomials \citep{PHL2017}. The two lowest-order terms in the Laguerre series, and Grad's expansion (with 21 moments) are identical when the fluid-dynamical limit is taken. The reader may consult Chapter 4 of \citet{Bal88} for further details. Second, we reiterate that (\ref{Psexp}) does not require any knowledge of the exact form of the `viscosity' and the heat flux. Hence, our results are also valid when $\pi_s^{ij}$ represents the pressure anisotropy (as in the Grad formalism).

\subsection{Explicit momentum transfer rate for the lowest order transport theory}
Using the result,
\begin{equation}
\partial_{\bm{\Delta}_{ss'}}\cdot\partial_{\bm{\Delta}_{ss'}}\Phi(\Delta_{ss'})=-4\pi {\cal N}_{1/2}(\bm{\Delta}_{ss'}),
\end{equation}
we can obtain the expression for the collisional momentum-transfer rate that includes the contributions from heat flux and viscosity,
\begin{widetext}
\begin{multline}\label{Fexact}
\bm{F}_{ss'}
=\frac{\bar{c}_{ss'}\mu_{ss'}}{2\Sigma_{ss'}^2}\partial_{\bm{\Delta}_{ss'}}\Biggr[\Phi(\Delta_{ss'})
+\left(\frac{\sigma_s^2}{4\Sigma_{ss'}^2}\frac{\bm{\pi}^{[0]}_s}{p_s}+\frac{\sigma_{s'}^2}{4\Sigma_{ss'}^2}\frac{\bm{\pi}^{[0]}_{s'}}{p_{s'}}\right):\partial_{\bm{\Delta}_{ss'}}^{(2)}
 \Phi(\Delta_{ss'})
\\
+\frac{\sigma_s^2\sigma_{s'}^2}{16\Sigma_{ss'}^4}\left(\frac{\bm{\pi}^{[0]}_s}{p_s}:\partial_{\bm{\Delta}_{ss'}}^{(2)}\right)\left(\frac{\bm{\pi}^{[0]}_{s'}}{p_{s'}}:\partial_{\bm{\Delta}_{ss'}}^{(2)}\right)
 \Phi(\Delta_{ss'})
-\frac{\sqrt{2}\pi}{5}\left(\frac{\sigma_s^2}{\Sigma_{ss'}^{2}}\frac{\bm{q}^{[0]}_s}{p_s\Sigma_{ss'}}-\frac{\sigma_{s'}^2}{\Sigma_{ss'}^{2}}\frac{\bm{q}^{[0]}_{s'}}{p_{s'}\Sigma_{ss'}}\right)\cdot\partial_{\bm{\Delta}_{ss'}}
{\cal N}_{1/2}(\bm{\Delta}_{ss'})
\\
-\frac{\pi}{10\sqrt{2}}\frac{\sigma_{s'}^2\sigma_s^2}{\Sigma_{ss'}^4}\Biggr(\frac{\bm{q}_{s}^{[0]}\cdot\partial_{\bm{\Delta}_{ss'}}}{p_{s}\Sigma_{ss'}}\frac{\bm{\pi}^{[0]}_{s'}}{p_{s'}}-\frac{\bm{q}_{s'}^{[0]}\cdot\partial_{\bm{\Delta}_{ss'}}}{p_{s'}\Sigma_{ss'}}\frac{\bm{\pi}^{[0]}_s}{p_s}\Biggr):\partial_{\bm{\Delta}_{ss'}}^{(2)}{\cal N}_{1/2}(\bm{\Delta}_{ss'})
\\
+\frac{\pi}{50}\frac{\sigma_{s}^2\sigma_{s'}^2}{\Sigma_{ss'}^{4}}\frac{\bm{q}_{s}^{[0]}\cdot\partial_{\bm{\Delta}_{ss'}}}{p_{s}\Sigma_{ss'}}\frac{\bm{q}_{s'}^{[0]}\cdot\partial_{\bm{\Delta}_{ss'}}}{p_{s'}\Sigma_{ss'}}\partial_{\bm{\Delta}_{ss'}}\cdot\partial_{\bm{\Delta}_{ss'}}{\cal N}_{1/2}(\bm{\Delta}_{ss'})
\Biggr]
\end{multline}
\end{widetext}
This represents an exact result for the collisional momentum-transfer rate, since no assumptions were invoked thus far apart from our ansatz (\ref{Psexp}).

\subsection{Expression for the resistivity tensor}
Let us now specialize to the case of $\bm{F}_{ei}$ and invoke the
standard ordering described in Sec.~\ref{SecSummary}. We see that
\begin{equation}
\Sigma_{ei}\approx \sigma_e,
\end{equation}
which leads us to
\begin{equation}
\frac{\bar{c}_{ei} \mu_{ei}}{2\Sigma^2_{ei}}\approx \eta_0\,\frac{3\sqrt{\pi}}{2\sqrt{2}}e^2n^2\sigma_{e}.
\end{equation}
With our choice of ordering, we also find
\begin{equation}
\frac{\sigma_e^2\sigma_i^2}{\Sigma_{ei}^4}\approx\frac{\sigma_i^2}{\sigma_e^2}\approx\frac{m_e}{m_i}\ll 1, \qquad \frac{\sigma_i^2}{\sigma_e^2}\frac{p_e}{p_i}\approx \frac{m_e}{m_i}\ll 1.
\end{equation}
If we further assume that the ion viscosity is not orders of magnitude larger than the electron viscosity, we can approximate
\begin{equation} \label{ElVisc}
\frac{\sigma_e^2}{4\Sigma_{ei}^2}\frac{\bm{\pi}^{[0]}_e}{p_e}+\frac{\sigma_{i}^2}{4\Sigma_{ei}^2}\frac{\bm{\pi}^{[0]}_{i}}{p_{i}}
\approx\frac{1}{4p_e}\left(\bm{\pi}^{[0]}_e+\frac{m_e}{m_i}\bm{\pi}^{[0]}_{i}\right)\approx\frac{\bm{\pi}^{[0]}_e}{4p_e},
\end{equation}
but it must be noted that one could easily impose an alternative ordering, for e.g. dropping  $\bm{\pi}^{[0]}_e$. A similar assumption for the heat flux leads us to
\begin{equation}
\frac{\sigma_e^2}{\Sigma_{ei}^{2}}\frac{\bm{q}^{[0]}_e}{p_e\Sigma_{ei}}-\frac{\sigma_{i}^2}{\Sigma_{ei}^{2}}\frac{\bm{q}^{[0]}_{i}}{p_{i}\Sigma_{ei}}\approx \frac{1}{p_e\Sigma_{ei}}\left(\bm{q}^{[0]}_e-\frac{m_e}{m_i}\bm{q}^{[0]}_{i}\right)\approx\frac{\bm{q}^{[0]}_e}{p_e\sigma_e}.
\end{equation}
Finally, we observe that
\begin{equation}
\sqrt{2}ne\sigma_e\bm{\Delta}_{ei}=-\bm{J}, \qquad \Delta_{ei}=\sqrt{\frac{m_e}{ne^2}\frac{J^2}{2p_e}}.
\end{equation}
Thus, the collisional contributions to the generalized Ohm's law are expressible as
\begin{multline} \label{OhmIE}
\frac{\bm{F}_{ei}}{ne}
=\eta_0\Biggr[-\frac{3\sqrt{\pi}}{4}\frac{\Phi'(\Delta_{ei})}{\Delta_{ei}}\bm{J}
\\
-\frac{3\sqrt{\pi}}{16}\frac{1}{\Delta_{ei}}\left(\Phi''(\Delta_{ei})-3\frac{\Phi'(\Delta_{ei})}{\Delta_{ei}}\right)'\frac{\bm{\pi}^{[0]}_e}{p_e}:\frac{\bm{J}\bm{J}\bm{J}}{J^2}
\\
-\frac{3\sqrt{\pi}}{16}\frac{2}{\Delta_{ei}^2}\left(\Phi''(\Delta_{ei})-\frac{\Phi'(\Delta_{ei})}{\Delta_{ei}}\right)\frac{\bm{\pi}^{[0]}_e}{p_e}\cdot\bm{J}
\\
+\frac{3}{5} e^{-\Delta_{ei}^2}\frac{ne\,\bm{q}^{[0]}_e}{p_e}\cdot
\left(\mathbb{I}-2\Delta^2_{ei}\frac{\bm{J}\bm{J}}{J^2}\right)
\Biggr],
\end{multline}
and $\Phi'$ denotes differentiation of $\Phi$ with respect to $\Delta_{ei}$. In many real-world systems, $\Delta_{ei} \ll 1$, which allows us to carry out an expansion in this parameter. To first order, we find
\begin{multline} \label{FeiDelExp}
\frac{\bm{F}_{ei}}{ne}
=\eta_0\Biggr[\left(1-\frac{3\Delta_{ei}^2}{5}\right)\bm{J}
+\frac{3\Delta_{ei}^2}{7}\frac{\bm{\pi}^{[0]}_e}{p_e}:\frac{\bm{J}\bm{J}}{J^2}\bm{J}
\\-\frac{3}{5}\left(1-\frac{5\Delta_{ei}^2}{7}\right)\frac{\bm{\pi}^{[0]}_e}{p_e}\cdot\bm{J}
\\
+\frac{3}{5} \frac{ne\,\bm{q}^{[0]}_e}{p_e}\cdot
\left((1-\Delta_{ei}^2)\mathbb{I}-2\Delta^2_{ei}\frac{\bm{J}\bm{J}}{J^2}\right)
\Biggr].
\end{multline}
Hence, the lowest order contribution is given by the fairly simple expression
\begin{equation} \label{ResLowOrd}
\frac{\bm{F}_{ei}}{ne}
=\eta_0\left(\bm{J}
-\frac{3}{5}\frac{\bm{\pi}^{[0]}_e}{p_e}\cdot\bm{J}
+\frac{3}{5} \frac{ne\,\bm{q}^{[0]}_e}{p_e}
\right).
\end{equation}
Let us initially consider the first two terms on the RHS of (\ref{ResLowOrd}). The first term corresponds to the conventional Spitzer resistivity, while the second arises from the viscosity (or pressure anisotropy). Collectively, these two terms can be represented as an effective `resistivity tensor' that is independent of the current. However, including higher-order contributions makes it dependent on the current and other fluid variables, as seen from (\ref{FeiDelExp}).

An inspection of (\ref{ResLowOrd}) makes it apparent that the third term is unlike the first two - the latter duo vanish in the limit $\bm{J} \rightarrow 0$, but the former, arising from the heat flux, still survives. This term is responsible for giving rise to the thermal force in (\ref{eq:braginskii}). In fact, the above statement can be generalized to show that, in the limit of $\Delta_{ei} \rightarrow 0$, the $(2p+1)$ velocity-space moments of the distribution function (for $p > 0$) do not vanish when the final expression for $\bm{F}_{ei}$ is computed.

\subsection{Comparison with Braginskii's model}
As we have stated earlier, the expression for $\bm{F}_{ei}$ was independent of the exact form chosen for the heat flux (and viscosity). In the limit that the magnetic field dominates collisional effects, Braginskii's expression for the electron heat flux term reduces to
\begin{multline} \label{qeBrag}
\frac{ne\,\bm{q}_e}{p_e}=-0.71\frac{\bm{B}\bm{B}}{B^2}\cdot\bm{J}
\\-\frac{3.16}{\eta_0e}\frac{\bm{B}\bm{B}}{B^2}\cdot\nabla T_e
-\frac{5}{2}\frac{n}{B}\frac{\bm{B}}{B}\times\nabla T_e .
\end{multline}
If we now make the assumption that the lowest-order electron heat flux $\bm{q}_e^{[0]}$ is close to (\ref{qeBrag}), we find that the expression for the collisional momentum transfer becomes
\begin{multline} \label{FeiBBrag}
\frac{\bm{F}_{ei}}{ne}=\eta_0\left[0.57\frac{\bm{B}\bm{B}}{B^2}+\left(\mathbb{I}-\frac{\bm{B}\bm{B}}{B^2}\right)-\frac{3}{5}\frac{\bm{\pi}^{[0]}_e}{p_e}\right]\cdot\bm{J}\\
-1.9\frac{\bm{B}\bm{B}}{e B^2}\cdot\nabla T_e-\frac{3\eta_0}{2}\frac{n}{B}\frac{\bm{B}}{B}\times\nabla T_e
\end{multline}
This is close to the expressions (\ref{eq:braginskii}) and (\ref{ThermFo}) obtained by Braginskii \citep{Brag65}, except for the factor of $\sim 2.5$ difference in the parallel thermal-force contribution. Note that (\ref{FeiBBrag}) also contains the electron `viscosity' term that is absent in (\ref{eq:braginskii}), but that has to do with our choice of ordering - instead, if we dropped the electron viscosity in (\ref{ElVisc}), it would not appear in (\ref{FeiBBrag}). 

Hence, to summarize, we find that our lowest-order result is in line with Braginskii's transport calculations, apart from the fact that our parallel thermal force is about $2.5$ times higher than Braginskii's. This discrepancy is related to the fact that Braginskii's calculation involved not only the lowest order Laguerre polynomials but also higher-order contributions. We wish to reiterate that our model can be easily extended to also include the higher-order transport coefficients, and that it also captures non-linear contributions, as demonstrated by (\ref{Fexact}), (\ref{OhmIE}) and (\ref{FeiDelExp}).

\section{The physical implications of the resistivity tensor} \label{SecMaxwell}
In this Section, we discuss some of the physical implications arising from the resistivity tensor, both with respect to the current nonlinearity and the pressure anisotropy. 

\subsection{The physical interpretation and significance of the parameter $\Delta_{ei}$} \label{DelPar}
Since the nonlinearities in the resistivity (\ref{OhmIE}) are introduced when the parameter $\Delta_{ei}\neq 0$, it is important to gauge its physical meaning. On dimensional grounds, it is clear that
\begin{equation} \label{DeltaPhys}
\Delta_{ei} \sim \sqrt{\frac{m_e B^2}{2 n \mu_0^2 e^2 L^2 p_e}} \sim \left(\frac{d_e}{L}\right) \beta_e^{-1/2},
\end{equation}
where we have used the sole assumption that the spatial gradient(s) can be replaced by the inverse scale length $1/L$. In the above relation, $d_e$ and $\beta_e$ denote the characteristic electron skin depth and electron plasma beta respectively. A large number of fusion \citep{KT73,HM92,Fre14}, space and astrophysical \citep{KR95,GP04,BG05,Kul05,Pri14} plasmas are characterized by low values of the plasma beta. However, the normalized electron skin depth $\left(d_e/L\right)$ is extremely small, implying that $\Delta_{ei} \ll 1$ is commonplace.

Yet, there are some crucial instances where $\Delta_{ei}$ can be $\mathcal{O}(1)$ or even higher. This typically occurs in phenomena where the relevant length scales are on the order of the electron skin depth. Earth \citep{KR95,Butal16}, planetary \citep{KR95,Gomb98} and pulsar \citep{HL06,PS14} magnetospheres, as well as solar \citep{SGRK,Alex09,BC13} and stellar \citep{Gomb98,LC99} winds represent particularly important astrophysical examples wherein $\Delta_{ei} \gtrsim \mathcal{O}(1)$ can occur. Sawtooth crashes in tokamaks also merit a mention in this regard \citep{Hast97}. In these cases, at internal boundary layers (realized, for example, during collisionless reconnection dynamics), the plasma current density can be extremely localized. For such systems, we emphasize that the conventional Spitzer value, for the electric resistivity, may not be accurate. Hence, our analysis highlights the important conceptual point that the resistivity must be specified with due caution for a given plasma.

We also note that the limit $\Delta_{ei} \gg 1$ corresponds to runaway solutions which are unphysical. Therefore, we will discuss the effects of $\Delta_{ei}$ using the expression~(\ref{FeiDelExp}) for the resistivity tensor, i.e. in the limit of $\Delta_{ei} \ll 1$. Even in this instance, there are subtleties involved as noted in \citet{SH53}. Naturally, the resistivity tensor (\ref{OhmIE}) would be valid as long as the Landau collision operator and the orderings are relevant.

\subsection{Implications of the current and pressure anisotropy dependence} \label{SSecCNRImp}

The second term on the RHS of (\ref{FeiDelExp}) represents a nonlinear current-dependent correction to the resistivity $\left(\mathcal{NR}\right)$ of the form
\begin{equation} \label{DeltaEffect}
\mathcal{NR} \sim \eta_0\Delta_{ei}^2 {J}; \quad \Delta_{ei} = \sqrt{\frac{m_e}{ne^2}\frac{J^2}{2p_e}}.
\end{equation}

A better understanding of the role of $\mathcal{NR}$ follows from carrying out a dimensional comparison of this term against the other terms appearing in the extended MHD Ohm's law. We shall begin by comparing (\ref{DeltaEffect}) with the electron inertia terms $\left(\mathcal{EI}\right)$. For the latter, we shall choose the 2nd or 3rd term appearing in the second line of (\ref{XMHDOhm}). Thus, we arrive at
\begin{equation} \label{EINRComp}
\frac{\mathcal{NR}}{\mathcal{EI}} \sim \frac{\eta_0 J^3/p_e}{|\nabla \cdot \left({\bf J}{\bf V}\right)|} \sim \left(\frac{\eta_0 J}{V B}\right) \left(\frac{B^2}{\mu_0 p_e}\right) \sim \mathrm{R}_m^{-1} \beta_e^{-1},
\end{equation}
where $\mathrm{R}_m$ is the characteristic magnetic Reynolds number and electron plasma beta respectively. Thus, if the conditions $\beta_e \ll 1$ and/or $\mathrm{R}_m \ll 1$ are satisfied \footnote{Although the limit of $\mathrm{R}_m \gg 1$ is widely prevalent in most plasma environments, the opposite limit also occurs in certain real world settings - see Chapter 5 of \citet{Dav01}). The early Riga dynamo experiments constitute one such interesting historical example \citep{Gail08}, and theoretical analyses of this dynamo regime are also existent \citep{Moff70,BS05}.}, the nonlinear resistivity could be dominant compared to the electron inertia terms. Next, a comparison of the magnitudes of $\mathcal{NR}$ and the Hall drift $\left(\mathcal{H}\right)$ leads us to
\begin{equation} \label{HallNRComp}
\frac{\mathcal{NR}}{\mathcal{H}} \sim \frac{m_e \eta_0 J^3/(e^2 n p_e)}{|{\bf J} \times {\bf B}|/(ne)} \sim \left(\mathrm{R}_m \beta_e\right)^{-1} \frac{1}{\tau \Omega_e},
\end{equation}
where $\tau$ is the characteristic timescale and $\Omega_e$ is the characteristic electron Larmor frequency.  The similarity between (\ref{HallNRComp}) and (\ref{EINRComp}) can be explained by expressing (\ref{HallNRComp}) in the following manner. 
\begin{equation}\label{HallEIRat}
\frac{\mathcal{NR}}{\mathcal{H}} =  \frac{\mathcal{NR}}{\mathcal{EI}} \times \frac{\mathcal{EI}}{\mathcal{H}},  
\end{equation}
and the first term on the RHS is identical to (\ref{EINRComp}). The second term on the RHS precisely equals $1/\left(\tau \Omega_e\right)$ \citep{KM14,LMT14,LMM16}, and can be greater than unity in certain scenarios. This regime, with dominant electron inertia effects, has often been investigated in collisionless reconnection studies \citep{OP93,BSD97,CGPPS}, and occasionally in dynamo theory \citep{Kle94}. 

A current-dependent resistivity is likely to be relevant in systems with thin and intense electron-scale current sheets. The ``anomalous'' resistivity ansatz $\eta_{\mathrm{eff}} = \eta_0 + \eta_1 J$ (for e.g., Chapter 14.8 of \citet{Kul05}) with $\eta_1 \geq 0$, which is commonly used, is assumed to arise due to contributions from micro-turbulence when the current density exceeds a stability threshold. The fact that the classical resistivity can, in some cases, be lower than the Spitzer value actually increases the magnitude of the anomaly that must be accounted for. Hence, these considerations are relevant for computer simulation codes which tend to sometimes use \emph{ad hoc} spatially localized resistivity for reasons of numerical stability. The current-dependent resistivity also plays an important role in facilitating the widely studied Petschek mechanism for fast magnetic reconnection \citep{Bisk86,BP07}. Other resistivity-dependent fast reconnection processes, mediated by shocks \citep{Kul05} or the plasmoid instability \citep{CLHB16}, may also be affected.

Let us now turn our attention to the second term on the RHS of (\ref{ResLowOrd}) that arises from the existence of pressure anisotropy (or viscosity, depending on the formalism invoked). The foremost effect of pressure anisotropy is that the resistivity becomes a (rank-2) tensor instead of being a scalar, which is along expected lines. Furthermore, an inspection of (\ref{ResLowOrd}) reveals that the inclusion of pressure anisotropy leads to a divergence from the classical Spitzer value, even in the limit of $\Delta_{ei} \rightarrow 0$. The extent of deviation from the Spitzer value is, not surprisingly, related to the degree of anisotropy. Thus, in systems with non-negligible agyrotropic contributions, the Spitzer resistivity would be inapplicable.

There have been several studies investigating the role of pressure tensor effects in governing magnetic reconnection and particle acceleration that have often entailed the use of fluid models \citep[e.g][]{KHW00,KHW01,HNSKZ,ELD13,LENK14,WHBG,NHH15,LDKE}. This is backed by considerable empirical evidence confirming the importance of electron pressure anisotropy in the magnetosheath, solar wind and laboratory experiments \citep[e.g][]{GLT05,JRY08,STetal08}. They demonstrated that the electron pressure anisotropy is decisive in setting the structure of the reconnection current sheet, and that it plays a crucial role in attaining fast magnetic reconnection rates. Thus, the inclusion of accurate pressure anisotropy contributions in fluid models is necessary for correctly evaluating important features of magnetic reconnection.

\section{Conclusion} \label{SecConc}
To sum up our results, we utilized certain convolution properties of the multidimensional Gauss-Hermite polynomials \citep{HLPC16,PHL2017} to obtain the nonlinear resistivity in the generalized Ohm's law, under a certain set of assumptions. The primary conclusions of our work can be summarized as follows.
\begin{itemize}
\item The collisional electron-ion momentum transfer rate is \emph{always} expressible as (\ref{eq:momentum-transfer}), and it exhibits a non-trivial dependence on all the fluid variables.
\item The lowest order contribution yielded the Spitzer resistivity, and the absence of any pressure anisotropy (or viscosity) and heat flux led to a nonlinear expression in the current that was identical to the standard result in Chapter 6 of \citet{KT73}. This limiting case(s) was useful in confirming the validity of our approach.
\item The Spitzer resistivity represents an overestimation of the actual collisional electron-ion momentum transfer rate in certain regimes, and should therefore be used with caution.
\item The introduction of the heat flux leads to the emergence of the thermal force in the collisional momentum transfer rate. If one further assumes that the heat flux consists of a flow component and a temperature gradient as in \citet{Brag65}, we showed that the expression for the collisional momentum-transfer rate is compatible with the latter model.
\item When pressure anisotropy (or viscosity) was introduced, the collisional momentum transfer rate could be effectively represented by a resistivity tensor which did not vanish in the limit $J \rightarrow 0$.
\item As the resistivity is naturally dependent on the fluid variables (such as the current, pressure anisotropy and heat flux), it may affect the behavior of magnetic reconnection processes with respect to what has been inferred in prior studies that relied upon the Spitzer resistivity.
\end{itemize}

Although many closure schemes for multi-moment fluid models (with pressure anisotropy and heat flux) have been developed and studied over the years \citep[e.g][]{SHD97,GPS05,Ram05,CB06}, many of them suffer from the limitation that they fail to take into account collisional effects, although a few notable exceptions do exist \citep[e.g][]{JH06,Ram07,MS16}. Clearly, the assumption of \emph{zero} collisions is artificial, and the field of galactic dynamics illustrates that there are key differences between weakly collisional and altogether collisionless (gravitational) plasmas \citep{BT08}. 

Thus, the formulation of multi-moment fluid models with self-consistent collisional contributions is very desirable. Moreover, if such contributions to the generalized Ohm's law can be cast into a form akin to that of the conventional resistivity, it is all the more beneficial since extant numerical codes need not be modified significantly. Our results represent a first step in this direction, and the next step will involve the formulation of the self-consistent 10-moment model \citep{PHL2017}. 

Hence, on account of the above reasons, we believe that the nonlinear contributions to the resistivity (dependent on the fluid variables) presented in this paper can serve as a stepping stone for more detailed investigations in the future. We end this paper by emphasizing that our treatment was purely classical in nature, i.e. it does not include neoclassical effects, turbulence, or other anomalous contributions, which can be larger than the results obtained herein.\\

\acknowledgments
The authors were supported by the Department of Energy Contract No. DE-AC02-09CH11466 and the National Science Foundation Grant Nos. AGS-1338944 and AGS-1552142 during the course of this work.


\begin{thebibliography}{61}%
\makeatletter
\providecommand \@ifxundefined [1]{%
 \@ifx{#1\undefined}
}%
\providecommand \@ifnum [1]{%
 \ifnum #1\expandafter \@firstoftwo
 \else \expandafter \@secondoftwo
 \fi
}%
\providecommand \@ifx [1]{%
 \ifx #1\expandafter \@firstoftwo
 \else \expandafter \@secondoftwo
 \fi
}%
\providecommand \natexlab [1]{#1}%
\providecommand \enquote  [1]{``#1''}%
\providecommand \bibnamefont  [1]{#1}%
\providecommand \bibfnamefont [1]{#1}%
\providecommand \citenamefont [1]{#1}%
\providecommand \href@noop [0]{\@secondoftwo}%
\providecommand \href [0]{\begingroup \@sanitize@url \@href}%
\providecommand \@href[1]{\@@startlink{#1}\@@href}%
\providecommand \@@href[1]{\endgroup#1\@@endlink}%
\providecommand \@sanitize@url [0]{\catcode `\\12\catcode `\$12\catcode
  `\&12\catcode `\#12\catcode `\^12\catcode `\_12\catcode `\%12\relax}%
\providecommand \@@startlink[1]{}%
\providecommand \@@endlink[0]{}%
\providecommand \url  [0]{\begingroup\@sanitize@url \@url }%
\providecommand \@url [1]{\endgroup\@href {#1}{\urlprefix }}%
\providecommand \urlprefix  [0]{URL }%
\providecommand \Eprint [0]{\href }%
\providecommand \doibase [0]{http://dx.doi.org/}%
\providecommand \selectlanguage [0]{\@gobble}%
\providecommand \bibinfo  [0]{\@secondoftwo}%
\providecommand \bibfield  [0]{\@secondoftwo}%
\providecommand \translation [1]{[#1]}%
\providecommand \BibitemOpen [0]{}%
\providecommand \bibitemStop [0]{}%
\providecommand \bibitemNoStop [0]{.\EOS\space}%
\providecommand \EOS [0]{\spacefactor3000\relax}%
\providecommand \BibitemShut  [1]{\csname bibitem#1\endcsname}%
\let\auto@bib@innerbib\@empty
\bibitem [{\citenamefont {{Spitzer}}(1956)}]{Spit56}%
  \BibitemOpen
  \bibfield  {author} {\bibinfo {author} {\bibfnamefont {L.}~\bibnamefont
  {{Spitzer}}},\ }\href@noop {} {\emph {\bibinfo {title} {{Physics of Fully
  Ionized Gases}}}}\ (\bibinfo  {publisher} {New York: Interscience
  Publishers},\ \bibinfo {year} {1956})\BibitemShut {NoStop}%
\bibitem [{\citenamefont {{Krall}}\ and\ \citenamefont
  {{Trivelpiece}}(1973)}]{KT73}%
  \BibitemOpen
  \bibfield  {author} {\bibinfo {author} {\bibfnamefont {N.~A.}\ \bibnamefont
  {{Krall}}}\ and\ \bibinfo {author} {\bibfnamefont {A.~W.}\ \bibnamefont
  {{Trivelpiece}}},\ }\href@noop {} {\emph {\bibinfo {title} {{Principles of
  plasma physics}}}},\ International Series in Pure and Applied Physics\
  (\bibinfo  {publisher} {McGraw-Hill},\ \bibinfo {year} {1973})\BibitemShut
  {NoStop}%
\bibitem [{\citenamefont {{Kivelson}}\ and\ \citenamefont
  {{Russell}}(1995)}]{KR95}%
  \BibitemOpen
  \bibfield  {author} {\bibinfo {author} {\bibfnamefont {M.~G.}\ \bibnamefont
  {{Kivelson}}}\ and\ \bibinfo {author} {\bibfnamefont {C.~T.}\ \bibnamefont
  {{Russell}}},\ }\href@noop {} {\emph {\bibinfo {title} {{Introduction to
  Space Physics}}}}\ (\bibinfo  {publisher} {Cambridge Univ. Press},\ \bibinfo
  {year} {1995})\BibitemShut {NoStop}%
\bibitem [{\citenamefont {{Hazeltine}}\ and\ \citenamefont
  {Waelbroeck}(2004)}]{HW04}%
  \BibitemOpen
  \bibfield  {author} {\bibinfo {author} {\bibfnamefont {R.~D.}\ \bibnamefont
  {{Hazeltine}}}\ and\ \bibinfo {author} {\bibfnamefont {F.~L.}\ \bibnamefont
  {Waelbroeck}},\ }\href@noop {} {\emph {\bibinfo {title} {The Framework Of
  Plasma Physics}}},\ Frontiers in Physics\ (\bibinfo  {publisher} {Westview
  Press},\ \bibinfo {year} {2004})\BibitemShut {NoStop}%
\bibitem [{\citenamefont {{Goedbloed}}\ and\ \citenamefont
  {{Poedts}}(2004)}]{GP04}%
  \BibitemOpen
  \bibfield  {author} {\bibinfo {author} {\bibfnamefont {J.~P.~H.}\
  \bibnamefont {{Goedbloed}}}\ and\ \bibinfo {author} {\bibfnamefont
  {S.}~\bibnamefont {{Poedts}}},\ }\href@noop {} {\emph {\bibinfo {title}
  {{Principles of Magnetohydrodynamics}}}}\ (\bibinfo  {publisher} {Cambridge
  Univ. Press},\ \bibinfo {year} {2004})\BibitemShut {NoStop}%
\bibitem [{\citenamefont {{Freidberg}}(2014)}]{Fre14}%
  \BibitemOpen
  \bibfield  {author} {\bibinfo {author} {\bibfnamefont {J.~P.}\ \bibnamefont
  {{Freidberg}}},\ }\href@noop {} {\emph {\bibinfo {title} {{Ideal MHD}}}}\
  (\bibinfo  {publisher} {Cambridge Univ. Press},\ \bibinfo {year}
  {2014})\BibitemShut {NoStop}%
\bibitem [{\citenamefont {{Dungey}}(1958)}]{Dung58}%
  \BibitemOpen
  \bibfield  {author} {\bibinfo {author} {\bibfnamefont {J.~W.}\ \bibnamefont
  {{Dungey}}},\ }\href@noop {} {\emph {\bibinfo {title} {{Cosmic
  Electrodynamics}}}}\ (\bibinfo  {publisher} {Cambridge Univ. Press},\
  \bibinfo {year} {1958})\BibitemShut {NoStop}%
\bibitem [{\citenamefont {{Gombosi}}(1998)}]{Gomb98}%
  \BibitemOpen
  \bibfield  {author} {\bibinfo {author} {\bibfnamefont {T.~I.}\ \bibnamefont
  {{Gombosi}}},\ }\href@noop {} {\emph {\bibinfo {title} {{Physics of the Space
  Environment}}}},\ Atmospheric and Space Science Series\ (\bibinfo
  {publisher} {Cambridge Univ. Press},\ \bibinfo {year} {1998})\BibitemShut
  {NoStop}%
\bibitem [{\citenamefont {{Gurnett}}\ and\ \citenamefont
  {{Bhattacharjee}}(2005)}]{BG05}%
  \BibitemOpen
  \bibfield  {author} {\bibinfo {author} {\bibfnamefont {D.~A.}\ \bibnamefont
  {{Gurnett}}}\ and\ \bibinfo {author} {\bibfnamefont {A.}~\bibnamefont
  {{Bhattacharjee}}},\ }\href@noop {} {\emph {\bibinfo {title} {{Introduction
  to Plasma Physics}}}}\ (\bibinfo  {publisher} {Cambridge Univ. Press},\
  \bibinfo {year} {2005})\BibitemShut {NoStop}%
\bibitem [{\citenamefont {{Kulsrud}}(2005)}]{Kul05}%
  \BibitemOpen
  \bibfield  {author} {\bibinfo {author} {\bibfnamefont {R.~M.}\ \bibnamefont
  {{Kulsrud}}},\ }\href@noop {} {\emph {\bibinfo {title} {{Plasma physics for
  astrophysics}}}},\ Princeton series in astrophysics\ (\bibinfo  {publisher}
  {Princeton Univ. Press},\ \bibinfo {year} {2005})\BibitemShut {NoStop}%
\bibitem [{\citenamefont {{Priest}}(2014)}]{Pri14}%
  \BibitemOpen
  \bibfield  {author} {\bibinfo {author} {\bibfnamefont {E.}~\bibnamefont
  {{Priest}}},\ }\href@noop {} {\emph {\bibinfo {title} {{Magnetohydrodynamics
  of the Sun}}}}\ (\bibinfo  {publisher} {Cambridge Univ. Press},\ \bibinfo
  {year} {2014})\BibitemShut {NoStop}%
\bibitem [{\citenamefont {{Hirvijoki}}\ \emph {et~al.}(2016)\citenamefont
  {{Hirvijoki}}, \citenamefont {{Lingam}}, \citenamefont {{Pfefferl{\'e}}},
  \citenamefont {{Comisso}}, \citenamefont {{Candy}},\ and\ \citenamefont
  {{Bhattacharjee}}}]{HLPC16}%
  \BibitemOpen
  \bibfield  {author} {\bibinfo {author} {\bibfnamefont {E.}~\bibnamefont
  {{Hirvijoki}}}, \bibinfo {author} {\bibfnamefont {M.}~\bibnamefont
  {{Lingam}}}, \bibinfo {author} {\bibfnamefont {D.}~\bibnamefont
  {{Pfefferl{\'e}}}}, \bibinfo {author} {\bibfnamefont {L.}~\bibnamefont
  {{Comisso}}}, \bibinfo {author} {\bibfnamefont {J.}~\bibnamefont {{Candy}}},
  \ and\ \bibinfo {author} {\bibfnamefont {A.}~\bibnamefont
  {{Bhattacharjee}}},\ }\href {\doibase 10.1063/1.4960669} {\bibfield
  {journal} {\bibinfo  {journal} {Phys. Plasmas}\ }\textbf {\bibinfo {volume}
  {23}},\ \bibinfo {eid} {080701} (\bibinfo {year} {2016})}\BibitemShut
  {NoStop}%
\bibitem [{\citenamefont {{Pfefferl{\'e}}}\ \emph {et~al.}(2017)\citenamefont
  {{Pfefferl{\'e}}}, \citenamefont {{Hirvijoki}},\ and\ \citenamefont
  {{Lingam}}}]{PHL2017}%
  \BibitemOpen
  \bibfield  {author} {\bibinfo {author} {\bibfnamefont {D.}~\bibnamefont
  {{Pfefferl{\'e}}}}, \bibinfo {author} {\bibfnamefont {E.}~\bibnamefont
  {{Hirvijoki}}}, \ and\ \bibinfo {author} {\bibfnamefont {M.}~\bibnamefont
  {{Lingam}}},\ }\href@noop {} {\bibfield  {journal} {\bibinfo  {journal}
  {submitted to Phys. Plasmas}\ } (\bibinfo {year} {2017})},\ \Eprint
  {http://arxiv.org/abs/1701.08037} {arXiv:1701.08037 [physics.plasm-ph]}
  \BibitemShut {NoStop}%
\bibitem [{\citenamefont {{Fitzpatrick}}(2014)}]{Fitz14}%
  \BibitemOpen
  \bibfield  {author} {\bibinfo {author} {\bibfnamefont {R.}~\bibnamefont
  {{Fitzpatrick}}},\ }\href@noop {} {\emph {\bibinfo {title} {Plasma Physics:
  An Introduction}}}\ (\bibinfo  {publisher} {CRC Press},\ \bibinfo {year}
  {2014})\BibitemShut {NoStop}%
\bibitem [{\citenamefont {{Hazeltine}}\ and\ \citenamefont
  {{Meiss}}(1992)}]{HM92}%
  \BibitemOpen
  \bibfield  {author} {\bibinfo {author} {\bibfnamefont {R.~D.}\ \bibnamefont
  {{Hazeltine}}}\ and\ \bibinfo {author} {\bibfnamefont {J.~D.}\ \bibnamefont
  {{Meiss}}},\ }\href@noop {} {\emph {\bibinfo {title} {Plasma Confinement}}},\
  \bibinfo {series} {Frontiers in Physics}, Vol.~\bibinfo {volume} {86}\
  (\bibinfo  {publisher} {Addison-Wesley},\ \bibinfo {year} {1992})\BibitemShut
  {NoStop}%
\bibitem [{\citenamefont {{Braginskii}}(1965)}]{Brag65}%
  \BibitemOpen
  \bibfield  {author} {\bibinfo {author} {\bibfnamefont {S.~I.}\ \bibnamefont
  {{Braginskii}}},\ }\href@noop {} {\bibfield  {journal} {\bibinfo  {journal}
  {Rev. Plasma Phys.}\ }\textbf {\bibinfo {volume} {1}},\ \bibinfo {pages}
  {205} (\bibinfo {year} {1965})}\BibitemShut {NoStop}%
\bibitem [{\citenamefont {{Chapman}}\ and\ \citenamefont
  {{Cowling}}(1970)}]{CC70}%
  \BibitemOpen
  \bibfield  {author} {\bibinfo {author} {\bibfnamefont {S.}~\bibnamefont
  {{Chapman}}}\ and\ \bibinfo {author} {\bibfnamefont {T.~G.}\ \bibnamefont
  {{Cowling}}},\ }\href@noop {} {\emph {\bibinfo {title} {{The mathematical
  theory of non-uniform gases}}}}\ (\bibinfo  {publisher} {Cambridge Univ.
  Press},\ \bibinfo {year} {1970})\BibitemShut {NoStop}%
\bibitem [{\citenamefont {{Balescu}}(1988)}]{Bal88}%
  \BibitemOpen
  \bibfield  {author} {\bibinfo {author} {\bibfnamefont {R.}~\bibnamefont
  {{Balescu}}},\ }\href@noop {} {\emph {\bibinfo {title} {Transport Processes
  in Plasmas}}}\ (\bibinfo  {publisher} {North-Holland},\ \bibinfo {year}
  {1988})\BibitemShut {NoStop}%
\bibitem [{\citenamefont {{Burch}}\ \emph {et~al.}(2016)\citenamefont
  {{Burch}}, \citenamefont {{Moore}}, \citenamefont {{Torbert}},\ and\
  \citenamefont {{Giles}}}]{Butal16}%
  \BibitemOpen
  \bibfield  {author} {\bibinfo {author} {\bibfnamefont {J.~L.}\ \bibnamefont
  {{Burch}}}, \bibinfo {author} {\bibfnamefont {T.~E.}\ \bibnamefont
  {{Moore}}}, \bibinfo {author} {\bibfnamefont {R.~B.}\ \bibnamefont
  {{Torbert}}}, \ and\ \bibinfo {author} {\bibfnamefont {B.~L.}\ \bibnamefont
  {{Giles}}},\ }\href {\doibase 10.1007/s11214-015-0164-9} {\bibfield
  {journal} {\bibinfo  {journal} {Space Sci. Rev.}\ }\textbf {\bibinfo {volume}
  {199}},\ \bibinfo {pages} {5} (\bibinfo {year} {2016})}\BibitemShut {NoStop}%
\bibitem [{\citenamefont {{Harding}}\ and\ \citenamefont {{Lai}}(2006)}]{HL06}%
  \BibitemOpen
  \bibfield  {author} {\bibinfo {author} {\bibfnamefont {A.~K.}\ \bibnamefont
  {{Harding}}}\ and\ \bibinfo {author} {\bibfnamefont {D.}~\bibnamefont
  {{Lai}}},\ }\href {\doibase 10.1088/0034-4885/69/9/R03} {\bibfield  {journal}
  {\bibinfo  {journal} {Rep. Prog. Phys.}\ }\textbf {\bibinfo {volume} {69}},\
  \bibinfo {pages} {2631} (\bibinfo {year} {2006})}\BibitemShut {NoStop}%
\bibitem [{\citenamefont {{Philippov}}\ and\ \citenamefont
  {{Spitkovsky}}(2014)}]{PS14}%
  \BibitemOpen
  \bibfield  {author} {\bibinfo {author} {\bibfnamefont {A.~A.}\ \bibnamefont
  {{Philippov}}}\ and\ \bibinfo {author} {\bibfnamefont {A.}~\bibnamefont
  {{Spitkovsky}}},\ }\href {\doibase 10.1088/2041-8205/785/2/L33} {\bibfield
  {journal} {\bibinfo  {journal} {Astrophys. J. Lett.}\ }\textbf {\bibinfo
  {volume} {785}},\ \bibinfo {eid} {L33} (\bibinfo {year} {2014})}\BibitemShut
  {NoStop}%
\bibitem [{\citenamefont {{Sahraoui}}\ \emph {et~al.}(2009)\citenamefont
  {{Sahraoui}}, \citenamefont {{Goldstein}}, \citenamefont {{Robert}},\ and\
  \citenamefont {{Khotyaintsev}}}]{SGRK}%
  \BibitemOpen
  \bibfield  {author} {\bibinfo {author} {\bibfnamefont {F.}~\bibnamefont
  {{Sahraoui}}}, \bibinfo {author} {\bibfnamefont {M.~L.}\ \bibnamefont
  {{Goldstein}}}, \bibinfo {author} {\bibfnamefont {P.}~\bibnamefont
  {{Robert}}}, \ and\ \bibinfo {author} {\bibfnamefont {Y.~V.}\ \bibnamefont
  {{Khotyaintsev}}},\ }\href {\doibase 10.1103/PhysRevLett.102.231102}
  {\bibfield  {journal} {\bibinfo  {journal} {Phys. Rev. Lett.}\ }\textbf
  {\bibinfo {volume} {102}},\ \bibinfo {eid} {231102} (\bibinfo {year}
  {2009})}\BibitemShut {NoStop}%
\bibitem [{\citenamefont {{Alexandrova}}\ \emph {et~al.}(2009)\citenamefont
  {{Alexandrova}}, \citenamefont {{Saur}}, \citenamefont {{Lacombe}},
  \citenamefont {{Mangeney}}, \citenamefont {{Mitchell}}, \citenamefont
  {{Schwartz}},\ and\ \citenamefont {{Robert}}}]{Alex09}%
  \BibitemOpen
  \bibfield  {author} {\bibinfo {author} {\bibfnamefont {O.}~\bibnamefont
  {{Alexandrova}}}, \bibinfo {author} {\bibfnamefont {J.}~\bibnamefont
  {{Saur}}}, \bibinfo {author} {\bibfnamefont {C.}~\bibnamefont {{Lacombe}}},
  \bibinfo {author} {\bibfnamefont {A.}~\bibnamefont {{Mangeney}}}, \bibinfo
  {author} {\bibfnamefont {J.}~\bibnamefont {{Mitchell}}}, \bibinfo {author}
  {\bibfnamefont {S.~J.}\ \bibnamefont {{Schwartz}}}, \ and\ \bibinfo {author}
  {\bibfnamefont {P.}~\bibnamefont {{Robert}}},\ }\href {\doibase
  10.1103/PhysRevLett.103.165003} {\bibfield  {journal} {\bibinfo  {journal}
  {Phys. Rev. Lett.}\ }\textbf {\bibinfo {volume} {103}},\ \bibinfo {eid}
  {165003} (\bibinfo {year} {2009})}\BibitemShut {NoStop}%
\bibitem [{\citenamefont {{Bruno}}\ and\ \citenamefont
  {{Carbone}}(2013)}]{BC13}%
  \BibitemOpen
  \bibfield  {author} {\bibinfo {author} {\bibfnamefont {R.}~\bibnamefont
  {{Bruno}}}\ and\ \bibinfo {author} {\bibfnamefont {V.}~\bibnamefont
  {{Carbone}}},\ }\href {\doibase 10.12942/lrsp-2013-2} {\bibfield  {journal}
  {\bibinfo  {journal} {Living Rev. Sol. Phys.}\ }\textbf {\bibinfo {volume}
  {10}},\ \bibinfo {pages} {2} (\bibinfo {year} {2013})}\BibitemShut {NoStop}%
\bibitem [{\citenamefont {{Lamers}}\ and\ \citenamefont
  {{Cassinelli}}(1999)}]{LC99}%
  \BibitemOpen
  \bibfield  {author} {\bibinfo {author} {\bibfnamefont {H.~J.~G.~L.~M.}\
  \bibnamefont {{Lamers}}}\ and\ \bibinfo {author} {\bibfnamefont {J.~P.}\
  \bibnamefont {{Cassinelli}}},\ }\href@noop {} {\emph {\bibinfo {title}
  {{Introduction to Stellar Winds}}}}\ (\bibinfo  {publisher} {Cambridge Univ.
  Press},\ \bibinfo {year} {1999})\BibitemShut {NoStop}%
\bibitem [{\citenamefont {{Hastie}}(1997)}]{Hast97}%
  \BibitemOpen
  \bibfield  {author} {\bibinfo {author} {\bibfnamefont {R.~J.}\ \bibnamefont
  {{Hastie}}},\ }\href@noop {} {\bibfield  {journal} {\bibinfo  {journal}
  {Astrophys. Space Sci.}\ }\textbf {\bibinfo {volume} {256}},\ \bibinfo
  {pages} {177} (\bibinfo {year} {1997})}\BibitemShut {NoStop}%
\bibitem [{\citenamefont {{Spitzer}}\ and\ \citenamefont
  {{H{\"a}rm}}(1953)}]{SH53}%
  \BibitemOpen
  \bibfield  {author} {\bibinfo {author} {\bibfnamefont {L.}~\bibnamefont
  {{Spitzer}}}\ and\ \bibinfo {author} {\bibfnamefont {R.}~\bibnamefont
  {{H{\"a}rm}}},\ }\href {\doibase 10.1103/PhysRev.89.977} {\bibfield
  {journal} {\bibinfo  {journal} {Phys. Rev.}\ }\textbf {\bibinfo {volume}
  {89}},\ \bibinfo {pages} {977} (\bibinfo {year} {1953})}\BibitemShut
  {NoStop}%
\bibitem [{Note1()}]{Note1}%
  \BibitemOpen
  \bibinfo {note} {Although the limit of $\protect \mathrm {R}_m \gg 1$ is
  widely prevalent in most plasma environments, the opposite limit also occurs
  in certain real world settings - see Chapter 5 of \protect \citet {Dav01}).
  The early Riga dynamo experiments constitute one such interesting historical
  example \protect \citep {Gail08}, and theoretical analyses of this dynamo
  regime are also existent \protect \citep {Moff70,BS05}.}\BibitemShut {Stop}%
\bibitem [{\citenamefont {{Kimura}}\ and\ \citenamefont
  {{Morrison}}(2014)}]{KM14}%
  \BibitemOpen
  \bibfield  {author} {\bibinfo {author} {\bibfnamefont {K.}~\bibnamefont
  {{Kimura}}}\ and\ \bibinfo {author} {\bibfnamefont {P.~J.}\ \bibnamefont
  {{Morrison}}},\ }\href {\doibase 10.1063/1.4890955} {\bibfield  {journal}
  {\bibinfo  {journal} {Phys. Plasmas}\ }\textbf {\bibinfo {volume} {21}},\
  \bibinfo {eid} {082101} (\bibinfo {year} {2014})}\BibitemShut {NoStop}%
\bibitem [{\citenamefont {{Lingam}}\ \emph {et~al.}(2015)\citenamefont
  {{Lingam}}, \citenamefont {{Morrison}},\ and\ \citenamefont
  {{Tassi}}}]{LMT14}%
  \BibitemOpen
  \bibfield  {author} {\bibinfo {author} {\bibfnamefont {M.}~\bibnamefont
  {{Lingam}}}, \bibinfo {author} {\bibfnamefont {P.~J.}\ \bibnamefont
  {{Morrison}}}, \ and\ \bibinfo {author} {\bibfnamefont {E.}~\bibnamefont
  {{Tassi}}},\ }\href {\doibase 10.1016/j.physleta.2014.12.008} {\bibfield
  {journal} {\bibinfo  {journal} {Phys. Lett. A}\ }\textbf {\bibinfo {volume}
  {379}},\ \bibinfo {pages} {570} (\bibinfo {year} {2015})}\BibitemShut
  {NoStop}%
\bibitem [{\citenamefont {{Lingam}}\ \emph {et~al.}(2016)\citenamefont
  {{Lingam}}, \citenamefont {{Miloshevich}},\ and\ \citenamefont
  {{Morrison}}}]{LMM16}%
  \BibitemOpen
  \bibfield  {author} {\bibinfo {author} {\bibfnamefont {M.}~\bibnamefont
  {{Lingam}}}, \bibinfo {author} {\bibfnamefont {G.}~\bibnamefont
  {{Miloshevich}}}, \ and\ \bibinfo {author} {\bibfnamefont {P.~J.}\
  \bibnamefont {{Morrison}}},\ }\href {\doibase 10.1016/j.physleta.2016.05.024}
  {\bibfield  {journal} {\bibinfo  {journal} {Phys. Lett. A}\ }\textbf
  {\bibinfo {volume} {380}},\ \bibinfo {pages} {2400} (\bibinfo {year}
  {2016})}\BibitemShut {NoStop}%
\bibitem [{\citenamefont {{Ottaviani}}\ and\ \citenamefont
  {{Porcelli}}(1993)}]{OP93}%
  \BibitemOpen
  \bibfield  {author} {\bibinfo {author} {\bibfnamefont {M.}~\bibnamefont
  {{Ottaviani}}}\ and\ \bibinfo {author} {\bibfnamefont {F.}~\bibnamefont
  {{Porcelli}}},\ }\href {\doibase 10.1103/PhysRevLett.71.3802} {\bibfield
  {journal} {\bibinfo  {journal} {Phys. Rev. Lett.}\ }\textbf {\bibinfo
  {volume} {71}},\ \bibinfo {pages} {3802} (\bibinfo {year}
  {1993})}\BibitemShut {NoStop}%
\bibitem [{\citenamefont {{Biskamp}}\ \emph {et~al.}(1997)\citenamefont
  {{Biskamp}}, \citenamefont {{Schwarz}},\ and\ \citenamefont
  {{Drake}}}]{BSD97}%
  \BibitemOpen
  \bibfield  {author} {\bibinfo {author} {\bibfnamefont {D.}~\bibnamefont
  {{Biskamp}}}, \bibinfo {author} {\bibfnamefont {E.}~\bibnamefont
  {{Schwarz}}}, \ and\ \bibinfo {author} {\bibfnamefont {J.~F.}\ \bibnamefont
  {{Drake}}},\ }\href {\doibase 10.1063/1.872211} {\bibfield  {journal}
  {\bibinfo  {journal} {Phys. Plasmas}\ }\textbf {\bibinfo {volume} {4}},\
  \bibinfo {pages} {1002} (\bibinfo {year} {1997})}\BibitemShut {NoStop}%
\bibitem [{\citenamefont {{Cafaro}}\ \emph {et~al.}(1998)\citenamefont
  {{Cafaro}}, \citenamefont {{Grasso}}, \citenamefont {{Pegoraro}},
  \citenamefont {{Porcelli}},\ and\ \citenamefont {{Saluzzi}}}]{CGPPS}%
  \BibitemOpen
  \bibfield  {author} {\bibinfo {author} {\bibfnamefont {E.}~\bibnamefont
  {{Cafaro}}}, \bibinfo {author} {\bibfnamefont {D.}~\bibnamefont {{Grasso}}},
  \bibinfo {author} {\bibfnamefont {F.}~\bibnamefont {{Pegoraro}}}, \bibinfo
  {author} {\bibfnamefont {F.}~\bibnamefont {{Porcelli}}}, \ and\ \bibinfo
  {author} {\bibfnamefont {A.}~\bibnamefont {{Saluzzi}}},\ }\href {\doibase
  10.1103/PhysRevLett.80.4430} {\bibfield  {journal} {\bibinfo  {journal}
  {Phys. Rev. Lett.}\ }\textbf {\bibinfo {volume} {80}},\ \bibinfo {pages}
  {4430} (\bibinfo {year} {1998})}\BibitemShut {NoStop}%
\bibitem [{\citenamefont {{Kleva}}(1994)}]{Kle94}%
  \BibitemOpen
  \bibfield  {author} {\bibinfo {author} {\bibfnamefont {R.~G.}\ \bibnamefont
  {{Kleva}}},\ }\href {\doibase 10.1103/PhysRevLett.73.1509} {\bibfield
  {journal} {\bibinfo  {journal} {Phys. Rev. Lett.}\ }\textbf {\bibinfo
  {volume} {73}},\ \bibinfo {pages} {1509} (\bibinfo {year}
  {1994})}\BibitemShut {NoStop}%
\bibitem [{\citenamefont {{Biskamp}}(1986)}]{Bisk86}%
  \BibitemOpen
  \bibfield  {author} {\bibinfo {author} {\bibfnamefont {D.}~\bibnamefont
  {{Biskamp}}},\ }\href {\doibase 10.1063/1.865670} {\bibfield  {journal}
  {\bibinfo  {journal} {Phys. Fluids}\ }\textbf {\bibinfo {volume} {29}},\
  \bibinfo {pages} {1520} (\bibinfo {year} {1986})}\BibitemShut {NoStop}%
\bibitem [{\citenamefont {{Birn}}\ and\ \citenamefont {{Priest}}(2007)}]{BP07}%
  \BibitemOpen
  \bibfield  {author} {\bibinfo {author} {\bibfnamefont {J.}~\bibnamefont
  {{Birn}}}\ and\ \bibinfo {author} {\bibfnamefont {E.~R.}\ \bibnamefont
  {{Priest}}},\ }\href@noop {} {\emph {\bibinfo {title} {{Reconnection of
  Magnetic Fields}}}}\ (\bibinfo  {publisher} {Cambridge Univ. Press},\
  \bibinfo {year} {2007})\BibitemShut {NoStop}%
\bibitem [{\citenamefont {{Comisso}}\ \emph {et~al.}(2016)\citenamefont
  {{Comisso}}, \citenamefont {{Lingam}}, \citenamefont {{Huang}},\ and\
  \citenamefont {{Bhattacharjee}}}]{CLHB16}%
  \BibitemOpen
  \bibfield  {author} {\bibinfo {author} {\bibfnamefont {L.}~\bibnamefont
  {{Comisso}}}, \bibinfo {author} {\bibfnamefont {M.}~\bibnamefont {{Lingam}}},
  \bibinfo {author} {\bibfnamefont {Y.-M.}\ \bibnamefont {{Huang}}}, \ and\
  \bibinfo {author} {\bibfnamefont {A.}~\bibnamefont {{Bhattacharjee}}},\
  }\href {\doibase 10.1063/1.4964481} {\bibfield  {journal} {\bibinfo
  {journal} {Phys. Plasmas}\ }\textbf {\bibinfo {volume} {23}},\ \bibinfo {eid}
  {100702} (\bibinfo {year} {2016})}\BibitemShut {NoStop}%
\bibitem [{\citenamefont {{Kuznetsova}}\ \emph {et~al.}(2000)\citenamefont
  {{Kuznetsova}}, \citenamefont {{Hesse}},\ and\ \citenamefont
  {{Winske}}}]{KHW00}%
  \BibitemOpen
  \bibfield  {author} {\bibinfo {author} {\bibfnamefont {M.~M.}\ \bibnamefont
  {{Kuznetsova}}}, \bibinfo {author} {\bibfnamefont {M.}~\bibnamefont
  {{Hesse}}}, \ and\ \bibinfo {author} {\bibfnamefont {D.}~\bibnamefont
  {{Winske}}},\ }\href {\doibase 10.1029/1999JA900396} {\bibfield  {journal}
  {\bibinfo  {journal} {J. Geophys. Res.}\ }\textbf {\bibinfo {volume} {105}},\
  \bibinfo {pages} {7601} (\bibinfo {year} {2000})}\BibitemShut {NoStop}%
\bibitem [{\citenamefont {{Kuznetsova}}\ \emph {et~al.}(2001)\citenamefont
  {{Kuznetsova}}, \citenamefont {{Hesse}},\ and\ \citenamefont
  {{Winske}}}]{KHW01}%
  \BibitemOpen
  \bibfield  {author} {\bibinfo {author} {\bibfnamefont {M.~M.}\ \bibnamefont
  {{Kuznetsova}}}, \bibinfo {author} {\bibfnamefont {M.}~\bibnamefont
  {{Hesse}}}, \ and\ \bibinfo {author} {\bibfnamefont {D.}~\bibnamefont
  {{Winske}}},\ }\href {\doibase 10.1029/1999JA001003} {\bibfield  {journal}
  {\bibinfo  {journal} {J. Geophys. Res.}\ }\textbf {\bibinfo {volume} {106}},\
  \bibinfo {pages} {3799} (\bibinfo {year} {2001})}\BibitemShut {NoStop}%
\bibitem [{\citenamefont {{Hesse}}\ \emph {et~al.}(2011)\citenamefont
  {{Hesse}}, \citenamefont {{Neukirch}}, \citenamefont {{Schindler}},
  \citenamefont {{Kuznetsova}},\ and\ \citenamefont {{Zenitani}}}]{HNSKZ}%
  \BibitemOpen
  \bibfield  {author} {\bibinfo {author} {\bibfnamefont {M.}~\bibnamefont
  {{Hesse}}}, \bibinfo {author} {\bibfnamefont {T.}~\bibnamefont {{Neukirch}}},
  \bibinfo {author} {\bibfnamefont {K.}~\bibnamefont {{Schindler}}}, \bibinfo
  {author} {\bibfnamefont {M.}~\bibnamefont {{Kuznetsova}}}, \ and\ \bibinfo
  {author} {\bibfnamefont {S.}~\bibnamefont {{Zenitani}}},\ }\href {\doibase
  10.1007/s11214-010-9740-1} {\bibfield  {journal} {\bibinfo  {journal} {Space
  Sci. Rev.}\ }\textbf {\bibinfo {volume} {160}},\ \bibinfo {pages} {3}
  (\bibinfo {year} {2011})}\BibitemShut {NoStop}%
\bibitem [{\citenamefont {{Egedal}}\ \emph {et~al.}(2013)\citenamefont
  {{Egedal}}, \citenamefont {{Le}},\ and\ \citenamefont {{Daughton}}}]{ELD13}%
  \BibitemOpen
  \bibfield  {author} {\bibinfo {author} {\bibfnamefont {J.}~\bibnamefont
  {{Egedal}}}, \bibinfo {author} {\bibfnamefont {A.}~\bibnamefont {{Le}}}, \
  and\ \bibinfo {author} {\bibfnamefont {W.}~\bibnamefont {{Daughton}}},\
  }\href {\doibase 10.1063/1.4811092} {\bibfield  {journal} {\bibinfo
  {journal} {Phys. Plasmas}\ }\textbf {\bibinfo {volume} {20}},\ \bibinfo {eid}
  {061201} (\bibinfo {year} {2013})}\BibitemShut {NoStop}%
\bibitem [{\citenamefont {{Le}}\ \emph {et~al.}(2014)\citenamefont {{Le}},
  \citenamefont {{Egedal}}, \citenamefont {{Ng}}, \citenamefont {{Karimabadi}},
  \citenamefont {{Scudder}}, \citenamefont {{Roytershteyn}}, \citenamefont
  {{Daughton}},\ and\ \citenamefont {{Liu}}}]{LENK14}%
  \BibitemOpen
  \bibfield  {author} {\bibinfo {author} {\bibfnamefont {A.}~\bibnamefont
  {{Le}}}, \bibinfo {author} {\bibfnamefont {J.}~\bibnamefont {{Egedal}}},
  \bibinfo {author} {\bibfnamefont {J.}~\bibnamefont {{Ng}}}, \bibinfo {author}
  {\bibfnamefont {H.}~\bibnamefont {{Karimabadi}}}, \bibinfo {author}
  {\bibfnamefont {J.}~\bibnamefont {{Scudder}}}, \bibinfo {author}
  {\bibfnamefont {V.}~\bibnamefont {{Roytershteyn}}}, \bibinfo {author}
  {\bibfnamefont {W.}~\bibnamefont {{Daughton}}}, \ and\ \bibinfo {author}
  {\bibfnamefont {Y.-H.}\ \bibnamefont {{Liu}}},\ }\href {\doibase
  10.1063/1.4861871} {\bibfield  {journal} {\bibinfo  {journal} {Phys.
  Plasmas}\ }\textbf {\bibinfo {volume} {21}},\ \bibinfo {eid} {012103}
  (\bibinfo {year} {2014})}\BibitemShut {NoStop}%
\bibitem [{\citenamefont {{Wang}}\ \emph {et~al.}(2015)\citenamefont {{Wang}},
  \citenamefont {{Hakim}}, \citenamefont {{Bhattacharjee}},\ and\ \citenamefont
  {{Germaschewski}}}]{WHBG}%
  \BibitemOpen
  \bibfield  {author} {\bibinfo {author} {\bibfnamefont {L.}~\bibnamefont
  {{Wang}}}, \bibinfo {author} {\bibfnamefont {A.~H.}\ \bibnamefont {{Hakim}}},
  \bibinfo {author} {\bibfnamefont {A.}~\bibnamefont {{Bhattacharjee}}}, \ and\
  \bibinfo {author} {\bibfnamefont {K.}~\bibnamefont {{Germaschewski}}},\
  }\href {\doibase 10.1063/1.4906063} {\bibfield  {journal} {\bibinfo
  {journal} {Phys. Plasmas}\ }\textbf {\bibinfo {volume} {22}},\ \bibinfo {eid}
  {012108} (\bibinfo {year} {2015})}\BibitemShut {NoStop}%
\bibitem [{\citenamefont {{Ng}}\ \emph {et~al.}(2015)\citenamefont {{Ng}},
  \citenamefont {{Huang}}, \citenamefont {{Hakim}}, \citenamefont
  {{Bhattacharjee}}, \citenamefont {{Stanier}}, \citenamefont {{Daughton}},
  \citenamefont {{Wang}},\ and\ \citenamefont {{Germaschewski}}}]{NHH15}%
  \BibitemOpen
  \bibfield  {author} {\bibinfo {author} {\bibfnamefont {J.}~\bibnamefont
  {{Ng}}}, \bibinfo {author} {\bibfnamefont {Y.-M.}\ \bibnamefont {{Huang}}},
  \bibinfo {author} {\bibfnamefont {A.}~\bibnamefont {{Hakim}}}, \bibinfo
  {author} {\bibfnamefont {A.}~\bibnamefont {{Bhattacharjee}}}, \bibinfo
  {author} {\bibfnamefont {A.}~\bibnamefont {{Stanier}}}, \bibinfo {author}
  {\bibfnamefont {W.}~\bibnamefont {{Daughton}}}, \bibinfo {author}
  {\bibfnamefont {L.}~\bibnamefont {{Wang}}}, \ and\ \bibinfo {author}
  {\bibfnamefont {K.}~\bibnamefont {{Germaschewski}}},\ }\href {\doibase
  10.1063/1.4935302} {\bibfield  {journal} {\bibinfo  {journal} {Phys.
  Plasmas}\ }\textbf {\bibinfo {volume} {22}},\ \bibinfo {eid} {112104}
  (\bibinfo {year} {2015})}\BibitemShut {NoStop}%
\bibitem [{\citenamefont {{Le}}\ \emph {et~al.}(2016)\citenamefont {{Le}},
  \citenamefont {{Daughton}}, \citenamefont {{Karimabadi}},\ and\ \citenamefont
  {{Egedal}}}]{LDKE}%
  \BibitemOpen
  \bibfield  {author} {\bibinfo {author} {\bibfnamefont {A.}~\bibnamefont
  {{Le}}}, \bibinfo {author} {\bibfnamefont {W.}~\bibnamefont {{Daughton}}},
  \bibinfo {author} {\bibfnamefont {H.}~\bibnamefont {{Karimabadi}}}, \ and\
  \bibinfo {author} {\bibfnamefont {J.}~\bibnamefont {{Egedal}}},\ }\href
  {\doibase 10.1063/1.4943893} {\bibfield  {journal} {\bibinfo  {journal}
  {Phys. Plasmas}\ }\textbf {\bibinfo {volume} {23}},\ \bibinfo {eid} {032114}
  (\bibinfo {year} {2016})}\BibitemShut {NoStop}%
\bibitem [{\citenamefont {{Gary}}\ \emph {et~al.}(2005)\citenamefont {{Gary}},
  \citenamefont {{Lavraud}}, \citenamefont {{Thomsen}}, \citenamefont
  {{Lefebvre}},\ and\ \citenamefont {{Schwartz}}}]{GLT05}%
  \BibitemOpen
  \bibfield  {author} {\bibinfo {author} {\bibfnamefont {S.~P.}\ \bibnamefont
  {{Gary}}}, \bibinfo {author} {\bibfnamefont {B.}~\bibnamefont {{Lavraud}}},
  \bibinfo {author} {\bibfnamefont {M.~F.}\ \bibnamefont {{Thomsen}}}, \bibinfo
  {author} {\bibfnamefont {B.}~\bibnamefont {{Lefebvre}}}, \ and\ \bibinfo
  {author} {\bibfnamefont {S.~J.}\ \bibnamefont {{Schwartz}}},\ }\href
  {\doibase 10.1029/2005GL023234} {\bibfield  {journal} {\bibinfo  {journal}
  {Geophys. Res. Lett.}\ }\textbf {\bibinfo {volume} {32}},\ \bibinfo {eid}
  {L13109} (\bibinfo {year} {2005})}\BibitemShut {NoStop}%
\bibitem [{\citenamefont {{Ji}}\ \emph {et~al.}(2008)\citenamefont {{Ji}},
  \citenamefont {{Ren}}, \citenamefont {{Yamada}}, \citenamefont {{Dorfman}},
  \citenamefont {{Daughton}},\ and\ \citenamefont {{Gerhardt}}}]{JRY08}%
  \BibitemOpen
  \bibfield  {author} {\bibinfo {author} {\bibfnamefont {H.}~\bibnamefont
  {{Ji}}}, \bibinfo {author} {\bibfnamefont {Y.}~\bibnamefont {{Ren}}},
  \bibinfo {author} {\bibfnamefont {M.}~\bibnamefont {{Yamada}}}, \bibinfo
  {author} {\bibfnamefont {S.}~\bibnamefont {{Dorfman}}}, \bibinfo {author}
  {\bibfnamefont {W.}~\bibnamefont {{Daughton}}}, \ and\ \bibinfo {author}
  {\bibfnamefont {S.~P.}\ \bibnamefont {{Gerhardt}}},\ }\href {\doibase
  10.1029/2008GL034538} {\bibfield  {journal} {\bibinfo  {journal} {Geophys.
  Res. Lett.}\ }\textbf {\bibinfo {volume} {35}},\ \bibinfo {eid} {L13106}
  (\bibinfo {year} {2008})}\BibitemShut {NoStop}%
\bibitem [{\citenamefont {{{\v S}tver{\'a}k}}\ \emph
  {et~al.}(2008)\citenamefont {{{\v S}tver{\'a}k}}, \citenamefont
  {{Tr{\'a}vn{\'{\i}}{\v c}ek}}, \citenamefont {{Maksimovic}}, \citenamefont
  {{Marsch}}, \citenamefont {{Fazakerley}},\ and\ \citenamefont
  {{Scime}}}]{STetal08}%
  \BibitemOpen
  \bibfield  {author} {\bibinfo {author} {\bibfnamefont {{\v S}.}~\bibnamefont
  {{{\v S}tver{\'a}k}}}, \bibinfo {author} {\bibfnamefont {P.}~\bibnamefont
  {{Tr{\'a}vn{\'{\i}}{\v c}ek}}}, \bibinfo {author} {\bibfnamefont
  {M.}~\bibnamefont {{Maksimovic}}}, \bibinfo {author} {\bibfnamefont
  {E.}~\bibnamefont {{Marsch}}}, \bibinfo {author} {\bibfnamefont {A.~N.}\
  \bibnamefont {{Fazakerley}}}, \ and\ \bibinfo {author} {\bibfnamefont
  {E.~E.}\ \bibnamefont {{Scime}}},\ }\href {\doibase 10.1029/2007JA012733}
  {\bibfield  {journal} {\bibinfo  {journal} {Journal of Geophysical Research
  (Space Physics)}\ }\textbf {\bibinfo {volume} {113}},\ \bibinfo {eid}
  {A03103} (\bibinfo {year} {2008})}\BibitemShut {NoStop}%
\bibitem [{\citenamefont {{Snyder}}\ \emph {et~al.}(1997)\citenamefont
  {{Snyder}}, \citenamefont {{Hammett}},\ and\ \citenamefont
  {{Dorland}}}]{SHD97}%
  \BibitemOpen
  \bibfield  {author} {\bibinfo {author} {\bibfnamefont {P.~B.}\ \bibnamefont
  {{Snyder}}}, \bibinfo {author} {\bibfnamefont {G.~W.}\ \bibnamefont
  {{Hammett}}}, \ and\ \bibinfo {author} {\bibfnamefont {W.}~\bibnamefont
  {{Dorland}}},\ }\href {\doibase 10.1063/1.872517} {\bibfield  {journal}
  {\bibinfo  {journal} {Phys. Plasmas}\ }\textbf {\bibinfo {volume} {4}},\
  \bibinfo {pages} {3974} (\bibinfo {year} {1997})}\BibitemShut {NoStop}%
\bibitem [{\citenamefont {{Goswami}}\ \emph {et~al.}(2005)\citenamefont
  {{Goswami}}, \citenamefont {{Passot}},\ and\ \citenamefont
  {{Sulem}}}]{GPS05}%
  \BibitemOpen
  \bibfield  {author} {\bibinfo {author} {\bibfnamefont {P.}~\bibnamefont
  {{Goswami}}}, \bibinfo {author} {\bibfnamefont {T.}~\bibnamefont {{Passot}}},
  \ and\ \bibinfo {author} {\bibfnamefont {P.~L.}\ \bibnamefont {{Sulem}}},\
  }\href {\doibase 10.1063/1.2096582} {\bibfield  {journal} {\bibinfo
  {journal} {Phys. Plasmas}\ }\textbf {\bibinfo {volume} {12}},\ \bibinfo {eid}
  {102109} (\bibinfo {year} {2005})}\BibitemShut {NoStop}%
\bibitem [{\citenamefont {{Ramos}}(2005)}]{Ram05}%
  \BibitemOpen
  \bibfield  {author} {\bibinfo {author} {\bibfnamefont {J.~J.}\ \bibnamefont
  {{Ramos}}},\ }\href {\doibase 10.1063/1.1884128} {\bibfield  {journal}
  {\bibinfo  {journal} {Phys. Plasmas}\ }\textbf {\bibinfo {volume} {12}},\
  \bibinfo {eid} {052102} (\bibinfo {year} {2005})}\BibitemShut {NoStop}%
\bibitem [{\citenamefont {{Chust}}\ and\ \citenamefont
  {{Belmont}}(2006)}]{CB06}%
  \BibitemOpen
  \bibfield  {author} {\bibinfo {author} {\bibfnamefont {T.}~\bibnamefont
  {{Chust}}}\ and\ \bibinfo {author} {\bibfnamefont {G.}~\bibnamefont
  {{Belmont}}},\ }\href {\doibase 10.1063/1.2138568} {\bibfield  {journal}
  {\bibinfo  {journal} {Phys. Plasmas}\ }\textbf {\bibinfo {volume} {13}},\
  \bibinfo {eid} {012506} (\bibinfo {year} {2006})}\BibitemShut {NoStop}%
\bibitem [{\citenamefont {{Ji}}\ and\ \citenamefont {{Held}}(2006)}]{JH06}%
  \BibitemOpen
  \bibfield  {author} {\bibinfo {author} {\bibfnamefont {J.-Y.}\ \bibnamefont
  {{Ji}}}\ and\ \bibinfo {author} {\bibfnamefont {E.~D.}\ \bibnamefont
  {{Held}}},\ }\href {\doibase 10.1063/1.2356320} {\bibfield  {journal}
  {\bibinfo  {journal} {Phys. Plasmas}\ }\textbf {\bibinfo {volume} {13}},\
  \bibinfo {eid} {102103} (\bibinfo {year} {2006})}\BibitemShut {NoStop}%
\bibitem [{\citenamefont {{Ramos}}(2007)}]{Ram07}%
  \BibitemOpen
  \bibfield  {author} {\bibinfo {author} {\bibfnamefont {J.~J.}\ \bibnamefont
  {{Ramos}}},\ }\href {\doibase 10.1063/1.2717595} {\bibfield  {journal}
  {\bibinfo  {journal} {Phys. Plasmas}\ }\textbf {\bibinfo {volume} {14}},\
  \bibinfo {eid} {052506} (\bibinfo {year} {2007})}\BibitemShut {NoStop}%
\bibitem [{\citenamefont {{Miller}}\ and\ \citenamefont
  {{Shumlak}}(2016)}]{MS16}%
  \BibitemOpen
  \bibfield  {author} {\bibinfo {author} {\bibfnamefont {S.~T.}\ \bibnamefont
  {{Miller}}}\ and\ \bibinfo {author} {\bibfnamefont {U.}~\bibnamefont
  {{Shumlak}}},\ }\href {\doibase 10.1063/1.4960041} {\bibfield  {journal}
  {\bibinfo  {journal} {Phys. Plasmas}\ }\textbf {\bibinfo {volume} {23}},\
  \bibinfo {eid} {082303} (\bibinfo {year} {2016})}\BibitemShut {NoStop}%
\bibitem [{\citenamefont {{Binney}}\ and\ \citenamefont
  {{Tremaine}}(2008)}]{BT08}%
  \BibitemOpen
  \bibfield  {author} {\bibinfo {author} {\bibfnamefont {J.}~\bibnamefont
  {{Binney}}}\ and\ \bibinfo {author} {\bibfnamefont {S.}~\bibnamefont
  {{Tremaine}}},\ }\href@noop {} {\emph {\bibinfo {title} {{Galactic Dynamics:
  Second Edition}}}}\ (\bibinfo  {publisher} {Princeton Univ. Press},\ \bibinfo
  {year} {2008})\BibitemShut {NoStop}%
\bibitem [{\citenamefont {{Davidson}}(2001)}]{Dav01}%
  \BibitemOpen
  \bibfield  {author} {\bibinfo {author} {\bibfnamefont {P.~A.}\ \bibnamefont
  {{Davidson}}},\ }\href@noop {} {\emph {\bibinfo {title} {{An Introduction to
  Magnetohydrodynamics}}}},\ Cambridge Texts in Applied Mathematics\ (\bibinfo
  {publisher} {Cambridge Univ. Press},\ \bibinfo {year} {2001})\BibitemShut
  {NoStop}%
\bibitem [{\citenamefont {{Gailitis}}\ \emph {et~al.}(2008)\citenamefont
  {{Gailitis}}, \citenamefont {{Gerbeth}}, \citenamefont {{Gundrum}},
  \citenamefont {{Lielausis}}, \citenamefont {{Platacis}},\ and\ \citenamefont
  {{Stefani}}}]{Gail08}%
  \BibitemOpen
  \bibfield  {author} {\bibinfo {author} {\bibfnamefont {A.}~\bibnamefont
  {{Gailitis}}}, \bibinfo {author} {\bibfnamefont {G.}~\bibnamefont
  {{Gerbeth}}}, \bibinfo {author} {\bibfnamefont {T.}~\bibnamefont
  {{Gundrum}}}, \bibinfo {author} {\bibfnamefont {O.}~\bibnamefont
  {{Lielausis}}}, \bibinfo {author} {\bibfnamefont {E.}~\bibnamefont
  {{Platacis}}}, \ and\ \bibinfo {author} {\bibfnamefont {F.}~\bibnamefont
  {{Stefani}}},\ }\href {\doibase 10.1016/j.crhy.2008.07.004} {\bibfield
  {journal} {\bibinfo  {journal} {Comptes Rendus Physique}\ }\textbf {\bibinfo
  {volume} {9}},\ \bibinfo {pages} {721} (\bibinfo {year} {2008})}\BibitemShut
  {NoStop}%
\bibitem [{\citenamefont {{Moffatt}}(1970)}]{Moff70}%
  \BibitemOpen
  \bibfield  {author} {\bibinfo {author} {\bibfnamefont {H.~K.}\ \bibnamefont
  {{Moffatt}}},\ }\href {\doibase 10.1017/S002211207000068X} {\bibfield
  {journal} {\bibinfo  {journal} {J. Fluid Mech.}\ }\textbf {\bibinfo {volume}
  {41}},\ \bibinfo {pages} {435} (\bibinfo {year} {1970})}\BibitemShut
  {NoStop}%
\bibitem [{\citenamefont {{Brandenburg}}\ and\ \citenamefont
  {{Subramanian}}(2005)}]{BS05}%
  \BibitemOpen
  \bibfield  {author} {\bibinfo {author} {\bibfnamefont {A.}~\bibnamefont
  {{Brandenburg}}}\ and\ \bibinfo {author} {\bibfnamefont {K.}~\bibnamefont
  {{Subramanian}}},\ }\href {\doibase 10.1016/j.physrep.2005.06.005} {\bibfield
   {journal} {\bibinfo  {journal} {Phys. Rep.}\ }\textbf {\bibinfo {volume}
  {417}},\ \bibinfo {pages} {1} (\bibinfo {year} {2005})}\BibitemShut {NoStop}%
\end{thebibliography}

%

\end{document}